\documentclass[sigconf]{acmart}

\usepackage{tcolorbox}
\usepackage{tabularx}
\usepackage{multirow}
\usepackage{array}
\usepackage{ragged2e}
\usepackage{tabularray}

\newcolumntype{L}{>{\raggedright\arraybackslash}X}

\AtBeginDocument{%
  }

\renewcommand\footnotetextcopyrightpermission[1]{} 
\setcopyright{none}
\definecolor{green}{HTML}{44AA99}
\definecolor{yellow}{HTML}{DDCC77}
\definecolor{blue}{HTML}{88CCEE}
\definecolor{red}{HTML}{CC6677}
\definecolor{darkred}{HTML}{DC3220}

\begin{document}

\title{Analyzing political stances on Twitter in the lead-up to the 2024 U.S. election}

\author{Hazem Ibrahim}
\affiliation{%
 \institution{New York University Abu Dhabi}
 \city{Abu Dhabi}
 \country{UAE}
 }
 
 \author{Farhan Khan}

\affiliation{%
 \institution{New York University Abu Dhabi}
 \city{Abu Dhabi}
 \country{UAE}
 }
 
\author{Hend Alabdouli}
 \authornote{Authors contributed equally to this article.}
\affiliation{%
 \institution{New York University Abu Dhabi}
 \city{Abu Dhabi}
 \country{UAE}
 }
 
\author{Maryam Almatrooshi}
\authornotemark[1]
\affiliation{%
 \institution{New York University Abu Dhabi}
 \city{Abu Dhabi}
 \country{UAE}
 }
 
\author{Tran Nguyen}
\authornotemark[1]
\affiliation{%
 \institution{New York University Abu Dhabi}
 \city{Abu Dhabi}
 \country{UAE}
 }
\author{Talal Rahwan}
\authornote{Joint-corresponding authors; \{talal.rahwan@nyu.edu, yasir.zaki@nyu.edu\}.}
\affiliation{%
 \institution{New York University Abu Dhabi}
 \city{Abu Dhabi}
 \country{UAE}
 }
 
\author{Yasir Zaki}
\authornotemark[2]
\affiliation{%
 \institution{New York University Abu Dhabi}
 \city{Abu Dhabi}
 \country{UAE}
 }

\renewcommand{\shortauthors}{Ibrahim et al.}

\begin{abstract}
Social media platforms play a pivotal role in shaping public opinion and amplifying political discourse, particularly during elections. However, the same dynamics that foster democratic engagement can also exacerbate polarization. To better understand these challenges, here, we investigate the ideological positioning of tweets related to the 2024 U.S. Presidential Election. To this end, we analyze 1,235 tweets from key political figures and 63,322 replies, and classify ideological stances into Pro-Democrat, Anti-Republican, Pro-Republican, Anti-Democrat, and Neutral categories. Using a classification pipeline involving three large language models (LLMs)—GPT-4o, Gemini-Pro, and Claude-Opus—and validated by human annotators, we explore how ideological alignment varies between candidates and constituents. We find that Republican candidates author significantly more tweets in criticism of the Democratic party and its candidates than vice versa, but this relationship does not hold for replies to candidate tweets. Furthermore, we highlight shifts in public discourse observed during key political events. By shedding light on the ideological dynamics of online political interactions, these results provide insights for policymakers and platforms seeking to address polarization and foster healthier political dialogue.
\end{abstract}


\maketitle
\pagestyle{plain}

\section{Introduction}
Social media platforms have become pivotal mediums for political discourse, often shaping public opinion through their recommendation algorithms and serving as hubs for information exchange. With millions of users engaging daily, these platforms amplify voices across the political spectrum. However, the same features that make social media valuable for democratic participation also render it susceptible to the spread of misinformation, polarization, and manipulation. As these dynamics increasingly influence elections and policy debates, understanding how political messaging and interactions unfold on social media is essential for fostering informed civic engagement and safeguarding democratic processes. Within this broader context, detecting political ideology stance on social media holds particular importance for policymakers, researchers, and social media platforms. Stance detection—the ability to identify whether a given message supports, opposes, or remains neutral toward a given topic, concept, or group—provides valuable insights into the ideological divisions, shifting narratives, and public sentiment surrounding key issues. For policymakers, such insights can inform strategies to address polarization, identify misinformation, and gauge public support for initiatives. Moreover, by enabling a deeper understanding of online discourse, accurate stance detection can guide interventions to promote healthier digital ecosystems while ensuring diverse viewpoints are represented in public debates.

The political content published on different social media platforms has been extensively analyzed over the last decade, ranging from examinations of the homophily of recommendations~\cite{boutyline2017social}, radicalization pathways~\cite{ibrahim2023youtube}, and the prevalence of echo chambers~\cite{bode2012facebooking, cinelli2021echo}. Most recently, in the context of the 2024 U.S. election, Ye et al.\ ~\cite{ye2024auditing} examined political recommendations made on Twitter through a socket puppet driven experiment, finding a default right-leaning bias in content exposure, with reduced exposure to opposing viewpoints. To classify ideological positioning and stance specifically, work prior to the introduction of large language models (LLMs) has relied on access to user characteristics, such as their retweet behaviour~\cite{stefanov2020predicting}, or network interactions~\cite{aldayel2019your}, to classify stance. Yet, the introduction of LLMs has offered researchers an increasingly reliable method to scale text annotation tasks previously done by human annotators or through supervised or unsupervised machine learning models. As shown by Ziems et al.~\cite{ziems2024can}, through zero-shot prompting, LLMs can reliably identify text characteristics such as stance, ideology, misinformation, and humor, among others. Of these characteristics, stance detection was the characteristic most reliably classified by GPT-4. Wu et al.\ demonstrated GPT-3.5's ability to produce accurate classifications of the ideology of U.S. political figures via analyzing their public statements~\cite{wu2023large}. Lastly, as shown by Gilardi et al.~\cite{gilardi2023chatgpt}, LLMs can accurately measure the topic of tweets, and the framing used by the authors of tweets to organize the message in a narrative manner. Taken together, prior work suggests that accurate categorizations of political texts can be obtained through LLM-driven, human-validated annotation pipelines~\cite{tornberg2024large, linegar2023large}.

Against this background, here, we seek to understand the ideological positioning of political candidates and their constituents in the lead up to the 2024 U.S. election on Twitter. 
To this end, we analyze a dataset of 1,235 tweets made by major U.S. political figures from May 1st, 2024 to November 1st, 2024 on the social media platform Twitter (now X, but will be referred to as Twitter for the remainder of this article). Furthermore, we analyze the 63,322 replies made to the aforementioned political figure tweets to answer the following research questions. \textbf{RQ1: } How do political candidates from either side of the political aisle position their tweets from an ideological perspective? \textbf{ RQ2: } How are the replies made to political candidate tweets ideologically positioned? \textbf{ RQ3: } Is engagement with political tweets correlated with a certain ideological positioning? \textbf{ RQ4: } Did the ideological positioning of political tweets change in response to major political events?

\section{Methodology}
\textbf{Data:} The dataset used in this study, collected by Balasubramanian et al.\ ~\cite{balasubramanian2024public}, encompasses large-scale social media discourse on Twitter related to the 2024 U.S. Presidential Election. Tweets were collected from May 1st to November 1st, 2024, and includes approximately 27 million publicly available political tweets. Each tweet in the dataset is accompanied by metadata, including timestamps, and user engagement metrics (such as replies, retweets, likes, and views).

\textbf{Stance classification pipeline:} To classify tweets from prominent political figures, we began by isolating tweets in the dataset made by Joe Biden, Kamala Harris, and Tim Walz of the Democratic party, and Donald Trump and JD Vance of the Republican party. These figures were selected as projected nominees for their respective parties at the time the dataset was collected, with Biden being replaced by Harris following his decision not to participate in the election in July 2024. This resulted in a total of 1,235 tweets from the aforementioned candidates. We then gathered all replies made to these tweets, amounting to 63,322 replies.

For classifying reply tweets, we utilized three LLMs, namely GPT-4o, Gemini-Pro, and Claude-Opus. Here, we make a small alteration to standard stance detection classification tasks. While prior work has largely aimed to classify the stance of a given statement or tweet into one of three classes (support, against, or none), we further separate the first two of these classes to delineate the support or criticism of a given political party. Specifically, we classify a given tweet into one of the following five categories: ``Anti-Democrat'' (AD), ``Anti-Republican'' (AR), ``Pro-Democrat'' (PD), ``Pro-Republican'' (PR), and ``Neutral'' (N). This is done to identify differences in the framing of ideological alignment within tweets in support of the two political parties. Indeed, a tweet aligned with the Republican party, for instance, may be directly in support of the Republican party or its candidate (PR), or potentially indirectly in support of the Republican party through criticism of the Democratic party or its candidate (AD). Thus, to identify the ideological alignment of reply tweets, the models were given the following prompt:

\begin{quote}
    \small
    \textit{Given the following tweet made by} \textbf{[CANDIDATE]} \textit{who is a} \textbf{[CANDIDATE PARTY]}\textit{, and its reply, classify the reply into one of the categories.}
    \newline
    \newline
    \textit{Candidate Tweet:} \textbf{[CANDIDATE TWEET]}

    \textit{Reply Tweet:} \textbf{[REPLY TWEET]}
    \newline
    \newline
    \textit{Anti-Democrat, Anti-Republican, Pro-Democrat, Pro-Republican, Neutral}    
\end{quote}

For classifying candidate tweets, a similar approach was used, with the following prompt provided:

\begin{quote}
    \small
    \textit{Given the following tweet made by} \textbf{[CANDIDATE]} \textit{who is a} \textbf{[CANDIDATE PARTY]}\textit{, classify the tweet into one of the categories.}
    \newline
    \newline
    \textit{Candidate Tweet:} \textbf{[CANDIDATE TWEET]}
    \newline
    \newline
    \textit{Anti-Democrat, Anti-Republican, Pro-Democrat, Pro-Republican, Neutral}    
\end{quote}

Lastly, we classify a random sample of 1000 tweets daily in the two weeks surrounding major political events, namely the first presidential debate on June 27th, the supreme court ruling on July 1st, and Trump's attempted assassination on July 13th. This amounted to a set of 32,832 tweets. To do so, we used the following prompt:

\begin{quote}
    \small
    \textit{Given the following tweet, classify the tweet into one of the categories.} \textit{Tweet:} \textbf{[TWEET]}
    \newline
    \newline
    \textit{Anti-Democrat, Anti-Republican, Pro-Democrat, Pro-Republican, Neutral, Not political}    
\end{quote}

Here, we add the ``Not political'' class to filter out tweets that may have been incorrectly classified as political during the scraping process of the prior work from which we obtain the dataset, although this amounted to only 8.06\% of the 32,832 tweets.

In all classification tasks, the tweets were first classified using GPT-4o and Gemini-Pro, while Claude-Opus was used to classify tweets when consensus was not reached between GPT-4o and Gemini-Pro. When consensus (agreement between two of the three models) was still not reached, a human annotator provided the final classification. Annotated tweet data and reproduction code can be found at \url{https://github.com/comnetsAD/Twitter_Stance_Detection}. Please refer to ~\cite{balasubramanian2024public} for the full political tweets dataset.

\textbf{Validation:} The outputs of the aforementioned text-based classification prompts were validated by three independent coders. In the case of reply tweet classification, we verify the accuracy of the model by manually annotating a sample of 250 replies made to Biden tweets and 250 replies made to Trump tweets. In the case of candidate tweet classification, we similarly annotate all 128 Republican candidate tweets as well as 128 randomly selected Democrat candidate tweets. Given that Donald Trump only returned to Twitter on August 12th, 2024 following a year-long break from the platform, the number of tweets made by Republican candidates within the dataset (128) were significantly fewer than those authored by Democrat candidates (1,107). Lastly, we manually annotated a random sample of 500 tweets out of those used to quantify the impact of major political events.

\begin{table}[!tb]
\centering
\begin{tabular}{l|c|c|c}
  & \begin{tabular}[c]{@{}c@{}}Candidate\\Tweets\end{tabular} & \begin{tabular}[c]{@{}c@{}}Reply\\Tweets\end{tabular} & \begin{tabular}[c]{@{}c@{}}Event\\Tweets\end{tabular}  \\ \hline
Accuracy (\%) &  91.0  & 92.2 & 90.2\\
Macro F1-Score & 91.1   & 85.5 & 88.3\\
Rater-LLM agreement &  $0.89^{***}$  &    $0.88^{***}$ &    $0.86^{***}$\\
Inter-LLM agreement &  $0.74^{***}$  &  $0.80^{***}$ &  $0.61^{***}$\\
Inter-rater agreement & $0.87^{***}$   &  $0.92^{***}$ &  $0.79^{***}$\\ \hline
\end{tabular}
\caption{LLM accuracy, Macro F1-score, rater-AI agreement (Fleiss’s $\kappa$), inter-LLM agreement (Krippendorf's $\alpha$), and inter-rater agreement (Krippendorf's $\alpha$) for text classification tasks (*: $p < 0.05$, **: $p < 0.01$, ***: $p < 0.001$).}
\vspace{-3em}
\label{table:annotation}
\end{table}

As shown in Table~\ref{table:annotation}, we find that the accuracy of the LLM consensus approach, computed as the proportion of tweets in which the majority vote of the coders matches the majority vote of the LLMs, exceeds 90\% in all three classification tasks. Furthermore, Fleiss's $\kappa$ was computed between the rater majority vote and the LLM majority vote to measure rater-model agreement for all classification tasks, achieving a score $\geq 0.86$ in each task, indicating strong agreement. In each case, to verify both LLM and human annotation reliability, inter-LLM and inter-rater agreement was also computed, respectively, using Krippendorf's $\alpha$, with each classification tasks receiving an inter-LLM $\alpha \geq 0.61$, and an inter-rater $\alpha \geq 0.79$, indicating moderate to strong agreement. 

\begin{figure*}[htbp!]
    \centering
    \includegraphics[width=\linewidth]{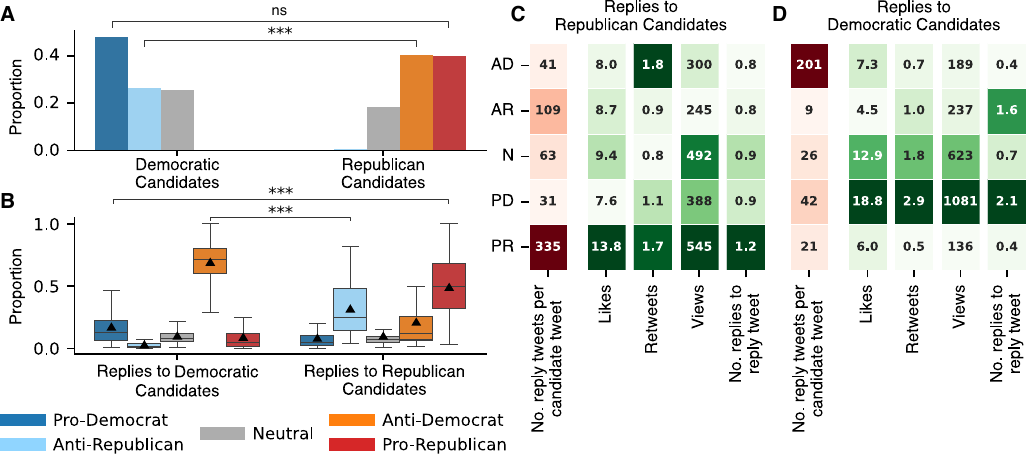}
    \caption{The proportion of candidate tweets (A) and candidate tweet replies (B) classified as Pro-Democrat, Anti-Republican, Neutral,  Anti-Democrat, or Pro-Republican (*: $p < 0.05$, **: $p < 0.01$, ***: $p < 0.001$). The average number of replies of a certain ideological stance made to (C) Democrat candidates and (D) Republican candidates per candidate tweet, as well as the average engagement metrics (likes, retweets, views, replies) such replies receive.}
    \Description{The proportion of candidate tweets (A) and candidate tweet replies (B) classified as Pro-Democrat (blue), Anti-Republican (light blue), Neutral (gray), Anti-Democrat (orange), or Pro-Republican (red). (*: $p < 0.05$, **: $p < 0.01$, ***: $p < 0.001$)}
    \label{fig:stance_rates}
\end{figure*}


\section{Results}

\textbf{RQ1:} We begin by analyzing how political candidates on either end of the political aisle frame their tweets from an ideological perspective. As can be seen in Figure~\ref{fig:stance_rates}A, of the 1,107 tweets made by Democrat political candidates, 48.1\% were framed in a manner such that they are in support of the Democratic party. In contrast, 26.4\% were positioned as criticisms of the Republican party, while a similar 25.5\% were neutrally positioned. On the other hand, of the 128 tweets made by Republican candidates, 40.2\% were positioned as in support of the Republican party, while 40.6\% were positioned as criticisms of the Democratic party. Statistical tests indicate that there were no significant differences with regards to the proportion of tweets in support of a given party's ideology (chi-squared test; $\chi^2 = 1.98$, $p = 0.16$), while there was a significant difference in the proportion of tweets framed as a criticism of the opposite party. Specifically, Republican candidates authored significantly more tweets in criticism of the Democratic party than vice-versa (chi-squared test; $\chi^2 = 11.55$, $p < 0.001$). 

\textbf{RQ2:} While candidate-authored tweets were largely positioned in support of the respective candidate's party (either pro-party or anti-opposite party), replies to such tweets did not exhibit the same pattern. As can be seen in Figure~\ref{fig:stance_rates}B, of the 36,254 replies to Democrat candidates, the majority were in opposition to the Democratic party (AD), with 69.2\% of replies being classified as such per candidate tweet on average. However, only 31.7\% of the 27,048 replies made to Republican candidate tweets were classified as AR on average, a significantly smaller proportion (two-sided independent t-test; $z = 15.19$, $p < 0.001$). In contrast, while 49.1\% of replies made to Republican candidates were PR, only 17.2\% of replies made to Democrat candidates were PD (two-sided independent t-test; $z = 12.60$, $p < 0.001$), suggesting that Republican-aligned Twitter users were more active in replying to the tweets of both Democrat and Republican candidates.

\textbf{RQ3:} Given that replies to political candidates on both sides of the political aisle skewed Republican (either Anti-Democrat or Pro-Republican), next, we analyze the engagement these replies received. Figures~\ref{fig:stance_rates}C and ~\ref{fig:stance_rates}D depict the average number of replies of a certain ideological stance made to Democrat candidates (Fig.~\ref{fig:stance_rates}C) and Republican candidates (Fig.~\ref{fig:stance_rates}D), as well as the average number of likes, retweets, views, and replies each reply tweet receives on average. Here, each column represents a separate heat map, or in other words, cell colors are computed relative to other values within the cell's column. Columns colored in red represent the average number of replies a candidate tweet receives on average, while those colored in green represent the various engagement metrics in question. As shown in Fig.~\ref{fig:stance_rates}C, for replies to Republican candidates, PR replies were both the most common type of reply (335 replies on average), and received the most engagement on average. However, for Democratic candidates, this is surprisingly not the case. As can be seen in Figure~\ref{fig:stance_rates}D, while the majority of replies received by Democratic candidates were classified as AD (201 replies on average), these replies received far less engagement than PD replies. This dichotomy in reply rate to reply engagement seen by Democratic candidates offers potential future inquiries into why this difference exists. Indeed, this difference could be driven by a number of factors, such as ``muting'' habits of supporters of the Democratic party, bot comments, or a result of algorithm-driven reply visibility.



\textbf{RQ4:}
\begin{figure*}[!ht]
        \centering
    \includegraphics[width=\linewidth]{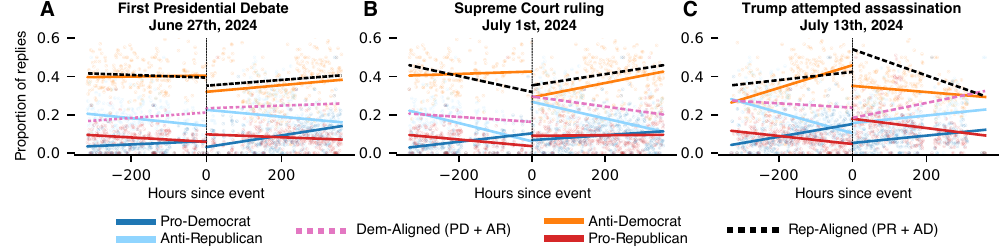}
    \vspace{-10pt}
    \caption{RDiT analysis of the ideological stance of tweets in the two weeks surrounding major political events. (A) First presidential debate between Biden and Trump; (B) Supreme Court ruling that a president has absolute immunity from criminal prosecution for core constitutional powers; (C) Trump's attempted assassination during a rally in Pennsylvania.}
    \Description{RDiT analysis of the ideological stance of tweets in the two weeks surrounding major events in the lead up to the U.S. election. (A) First presidential debate between Biden and Trump; (B) Supreme Court ruling that a president has absolute immunity from criminal prosecution for core constitutional powers; (C) Trump's attempted assassination during rally in Pennsylvania.}
    \label{fig:prepost}
\end{figure*}
To estimate the impact of major political events on the average ideological stance of tweets, we use a regression discontinuity in time (RDiT) design which is commonly used to study the treatment effect in quasi-experiments. Specifically, we focus on three major political events in the lead up to the 2024 U.S. election, namely, the first presidential debate between Biden and Trump on June 27th, the Supreme Court's ruling on presidential immunity on July 1st, and Trump's attempted assassination on July 13th. Specifically, here, we isolate 
a sample of 1000 tweets per day for each day two weeks before and after a given event.

Starting with the debate (Figure~\ref{fig:prepost}A), we estimate that this event did not have a statistically significant impact on Democrat-aligned ($p = 0.41$) or Republican-aligned ($p = 0.30$) tweets as a whole. However, further delineating that type of alignment within each set shows that the debate caused a 20.1\% drop in AD tweets ($p < 0.01$) coupled with an  66.0\% increase in PR tweets ($p < 0.01$). Conversely, the debate caused a sharp 58.8\% increase in AR tweets ($p < 0.001$) and a 49.4\% decrease in PD ($p < 0.05$) tweets, suggesting differing ideological reactions to the debate from supporters of each party. With regards to the Supreme Court ruling (Figure~\ref{fig:prepost}B), while again we do not find a significant impact on Republican aligned tweets as a whole ($p = 0.37$), we do see a significant increase in PR tweets (148\%, $p < 0.01$) and a significant drop in AD tweets (-31.6\%, $p < 0.001$). On the other hand, we do find a significant change in Democrat-aligned tweets (81\%, $p < 0.001$), primarily driven by a 336\% increase in AR tweets ($p < 0.001$), coupled with a 34\% drop in PD tweets ($p < 0.05$). Lastly, in response to Trump's attempted assassination, we find significant differences in both Democrat-aligned (-22\%, $p < 0.05$) and Republican-aligned tweets (27\%, $p < 0.001$) broadly. The drop in Democrat-aligned tweets was driven by 65\% ($p < 0.001$) drop in PD and a 51\% ($p < 0.001$) increase in AR tweets. In contrast, the rise in Republican-aligned tweets were primarily due to a 277\%  ($p < 0.001$) increase in PR tweets, with a 23\%  ($p < 0.001$)  drop in AD tweets. We repeat this analysis while isolating tweets made in the 3, 7, and 10 days before and after each event, and find broadly similar results.

Across all three events, we see two common patterns. Specifically, after each event, there was a significant increase in both AR and PR replies, coupled with significant decreases in both PD and AD replies, suggesting that discourse surrounding each event centered around support or criticism of Trump or the Republican party generally, from constituents of both parties. These results, as a whole, offer insights into the ideological positioning of tweets made both by political candidates specifically, and Twitter users broadly, in the lead-up to the 2024 U.S. election.

\vspace{-1em}

\section{Limitations and Future Work}
Due to the nature of the dataset analyzed, there are a number of limitations with regards to our analysis. Firstly, the dataset is not a comprehensive list of all political tweets on Twitter, and therefore, the results illustrated above only represent a sample of both candidate tweets and their replies. Nonetheless, given that our analysis focuses on a comparison in political alignment rates of candidate tweets and their replies, limitations based on tweet collection frequency should apply uniformly to tweets from both ends of the political spectrum. 

Future work may explore both the detection and influence of bot accounts in the context of promoting particular political ideologies. Furthermore, future work could examine the evolution of ideological positioning outside of the two-party system, incorporating third-party candidates and stances. Moreover, while our results aim to capture ideological positioning on Twitter, studies could look to collect and analyze such sentiments on other social media platforms, such as TikTok, BlueSky, or TruthSocial, each of which caters to a different age or political demographic.

\vspace{5pt}
\noindent \textbf{\underline{Ethics:}} All data analyzed in this study is publicly available. Informed consent was retrieved from the human annotators who validated model outputs.

\vspace{-0.5em}
\bibliographystyle{acm}
\bibliography{sample-base}

\begin{thebibliography}{10}

\bibitem{aldayel2019your}
{\sc Aldayel, A., and Magdy, W.}
\newblock Your stance is exposed! analysing possible factors for stance detection on social media.
\newblock {\em Proceedings of the ACM on Human-Computer Interaction 3}, CSCW (2019), 1--20.

\bibitem{balasubramanian2024public}
{\sc Balasubramanian, A., Zou, V., Narayana, H., You, C., Luceri, L., and Ferrara, E.}
\newblock A public dataset tracking social media discourse about the 2024 us presidential election on twitter/x.
\newblock {\em arXiv preprint arXiv:2411.00376\/} (2024).

\bibitem{bode2012facebooking}
{\sc Bode, L.}
\newblock Facebooking it to the polls: A study in online social networking and political behavior.
\newblock {\em Journal of Information Technology \& Politics 9}, 4 (2012).

\bibitem{boutyline2017social}
{\sc Boutyline, A., and Willer, R.}
\newblock The social structure of political echo chambers: Variation in ideological homophily in online networks.
\newblock {\em Political psychology\/} (2017).

\bibitem{cinelli2021echo}
{\sc Cinelli, M., De~Francisci~Morales, G., Galeazzi, A., Quattrociocchi, W., and Starnini, M.}
\newblock The echo chamber effect on social media.
\newblock {\em Proceedings of the National Academy of Sciences 118}, 9 (2021), e2023301118.

\bibitem{gilardi2023chatgpt}
{\sc Gilardi, F., Alizadeh, M., and Kubli, M.}
\newblock Chatgpt outperforms crowd workers for text-annotation tasks.
\newblock {\em PNAS 120}, 30 (2023), e2305016120.

\bibitem{ibrahim2023youtube}
{\sc Ibrahim, H., AlDahoul, N., Lee, S., Rahwan, T., and Zaki, Y.}
\newblock Youtube’s recommendation algorithm is left-leaning in the united states.
\newblock {\em PNAS nexus\/} (2023).

\bibitem{linegar2023large}
{\sc Linegar, M., Kocielnik, R., and Alvarez, R.~M.}
\newblock Large language models and political science.
\newblock {\em Frontiers in Political Science 5\/} (2023), 1257092.

\bibitem{stefanov2020predicting}
{\sc Stefanov, P., Darwish, K., Atanasov, A., and Nakov, P.}
\newblock Predicting the topical stance and political leaning of media using tweets.
\newblock In {\em Proceedings of the 58th Annual Meeting of the Association for Computational Linguistics\/} (2020).

\bibitem{tornberg2024large}
{\sc T{\"o}rnberg, P.}
\newblock Large language models outperform expert coders and supervised classifiers at annotating political social media messages.
\newblock {\em Social Science Computer Review\/} (2024), 08944393241286471.

\bibitem{wu2023large}
{\sc Wu, P.~Y., Nagler, J., Tucker, J.~A., and Messing, S.}
\newblock Large language models can be used to scale the ideologies of politicians in a zero-shot learning setting, april 2023.
\newblock {\em arXiv preprint arXiv:2303.12057\/} (2023).

\bibitem{ye2024auditing}
{\sc Ye, J., Luceri, L., and Ferrara, E.}
\newblock Auditing political exposure bias: Algorithmic amplification on twitter/x approaching the 2024 us presidential election.
\newblock {\em arXiv preprint arXiv:2411.01852\/} (2024).

\bibitem{ziems2024can}
{\sc Ziems, C., Held, W., Shaikh, O., Chen, J., Zhang, Z., and Yang, D.}
\newblock Can large language models transform computational social science?
\newblock {\em Computational Linguistics 50}, 1 (2024), 237--291.

\end{thebibliography}

\end{document}